\documentclass[aps,prb,twocolumn,showpacs,superscriptaddress]{revtex4}
\usepackage{graphicx} 
\begin{document}
\title
{\bf Decoherence and dephasing in multilevel systems interacting with thermal 
environment} 
\author{T. Hakio\u{g}lu and Kerim Savran}
\address{Department of Physics, Bilkent University, Bilkent, 06533
Ankara, Turkey}
\begin{abstract}
We examine the effect of multilevels on decoherence and dephasing properties 
of a quantum system consisting of a non-ideal two level subspace,  
identified as the qubit and 
a finite set of higher energy levels above this qubit subspace. The whole system 
is under interaction  
with an environmental bath through a Caldeira-Leggett type coupling. The model 
interaction we use can generate nonnegligible couplings between the qubit states 
and the higher levels upto $N\sim 10$.     
In contrast to the pure two-level system, in a multilevel system the quantum 
information may leak out of the qubit subspace through nonresonant as well as  
resonant excitations induced by  
 the environment. The decoherence properties of the qubit subspace is 
examined numerically using the master equation formalism of the system's reduced 
density matrix. We numerically examine the relaxation and dephasing times 
as the environmental frequency spectrum, the environmental temperature,   
and the multilevel system parameters are varied. We observe the 
influence of all energy scales in the noise spectrum on the short time dynamics  
implying the dominance of nonresonant transitions at short times. The  
relaxation and dephasing times calculated, strongly depend on $N$ for 
$4< N<10$ and saturate for $10 <N$. We also 
examine double degenerate systems with $4 \le N$ 
and observe a strong suppression (almost by two orders of magnitude) of   
the low temperature relaxation and dephasing rates. 

An important observation for $4 \le N$ in doubly degenerate energy 
configuration is that, we find a strong suppression of the RD rates 
for such systems relative to the singly degenerate ones. These results 
are also compared qualitatively with the relaxation rates found from the 
Fermi Golden Rule.
\end{abstract}
 
\maketitle  
\section{Introduction}
Currently a large number of model approaches are present for formulating 
the decoherence phenomena in the literature.  
 The original Caldeira-Leggett model\cite{CL} is based on a 
quantum system under the influence of a double well tunneling potential 
with a linear coupling to an infinite 
bath of harmonic oscillators. If the potential is {\it sufficiently} smooth  
and the high energy levels are sufficiently above the tunneling barrier this 
original model is normally represented as a two level system\cite{LG} (2LS) 
interacting with a bosonic 
environmental bath (spin-boson model). An incomplete list of this  
wide literature is provided in \cite{SB1,SB2,SBCS}. 
Another popular model of decoherence is the  
central spin system in which central 2LS couples to large number of 
environmental two level systems. The pros and cons of these two rival 
models have been extensively studied\cite{SBCS}. 

Realistically, and aside from the genuine 2LS, a large 
majority of physical systems suggested for qubit is far from being ideal. 
In a quantum computational environment, the parameters of the 
physical systems are manipulated to perform the gate operations. For instance, 
in a multilevel system (MLS), 
short time pulses used in the manipulation of the states in the qubit subspace 
 induce 
nonresonant transitions out of the qubit subspace. That nonresonant transitions 
contribute to the decoherence of the quantum system was recently addressed 
by Tian and Lloyd\cite{TL}. They suggest that after these transitions 
 are induced  
an optimized sequence of controlled pulses can be applied to cancel the 
nonresonant effects at arbitrary precision. The idea being physically correct,  
requires an additional fine knitting of error correction which undoubtedly  
 costs computational time. On the other hand, one may address the same issue  
 by seeking for an alternative solution: Can one understand the effect of the 
higher levels on decoherence in a well-parameterized MLS coupled to an 
environment? 

The MLS can itself be manifestly $N$-levelled or a truncated approximation  
of a larger system with much higher number of levels. Well known examples of 
 both cases
have been known. For the former, organic molecules with certain discrete
rotational symmetries and ground state low temperature configurations of  
single polimerized chains are good examples. The vibrational energy
spectra of atoms and molecules is a good example for the latter.

These type of realistic MLS can be found for instance 
in already well-examined superconducting systems such as the rf-SQUID in 
the charge, flux or phase regimes. We remark however, that a concise treatment 
of the decoherence effects in MLS has not been 
fully developed. This work is planned to be a modest step forward in that 
direction. 

In section II we give an introduction of the model MLS used in the  
present work. Here we merely concentrate on the properties of the environmentally 
induced {\it dipole} matrix elements. Section III recalls 
 the reduced density matrix (RDM) master equation formalism and adapts  
it for the coupling of the MLS to the environment. The noise correlator, which is 
considered to be in thermal equilibrium, and the system-noise kernel, for which no  
 Markovian assumption is made, are defined in section III.A. The results are 
presented together with the earlier observations of  
the 2LS (in section III.B) to allow a comparison with some of the  
established facts. The MLS with three or higher levels are examined in 
section III.C separately for $N=3$, 
$N=4$ and $4 < N$. The section III.C and the following section IV comprise most 
of the original results   
of the manuscript. In section IV, the relaxation times for MLS, which we 
produced numerically  in section III, are compared with the estimates found 
by using the semianalytic Fermi Golden Rule (FGR). 

\section{The Multilevel System}
In principle the majority of the well established 
methods (particularly the influence functional) used 
in the literature automatically accomodate multilevel dynamics. The results are 
generally hoped to be true for 2LS in the WKB limit at sufficiently low 
temperatures\cite{CL,LG}. Such techniques  
are also often preferable since they allow explicit analytic expressions for the 
decoherence times as functions of the system's parameters. On the other hand, 
exact methods are 
also available on pure 2LS\cite{SB2}. However, we need to accomodate in our 
parameters explicit degeneracy factors as well as the Hilbert space dimensions.   
 For the latter, the influence functional formalism has no parametric 
 dependence. In this work, in order to retain in our calculations 
the dependence on the system's Hilbert space dimension, we resort to the  
system's eigenenergy basis representation. 

\begin{figure}
\includegraphics[scale=0.45]{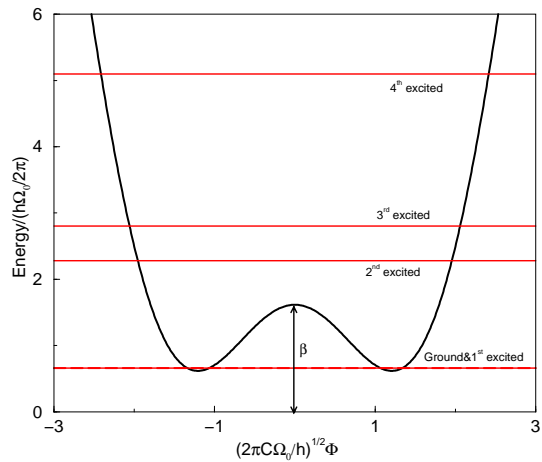}
\caption{The double-well potential and the eigenenergy configurations 
corresponding to the singly degenerate (SD) case. 
Here the numerical values of the dimensionless parameters for this SD 
 configuration are $\beta\simeq 1.616$ and $\gamma\simeq 1.753$. 
The harmonic energy scale and the degeneracy factors are respectively 
$\hbar\Omega_0=10^{-3} eV$ and $\eta_2=10^{6}$. 
Here the numerical values of the dimensionless parameters 
 are $\beta\simeq 1.616$ and $\gamma\simeq 1.753$.}  
\label{pot1}
\end{figure}

\begin{figure}
\includegraphics[scale=0.45]{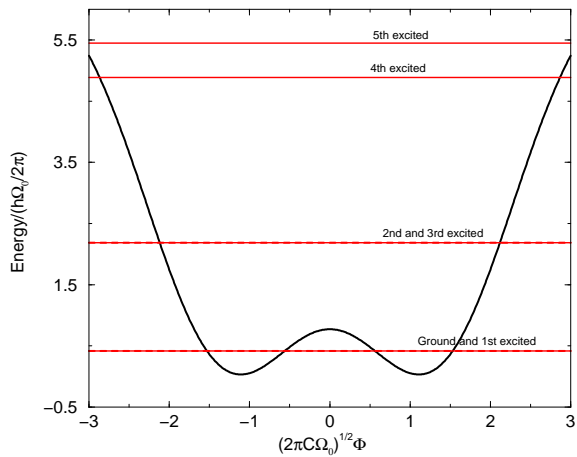}
\caption{The double-well potential and the eigenenergy configurations 
corresponding
to the doubly degenerate (DD) case considered throughout the manuscript.
The numerical values of the dimensionless parameters corresponding to this
DD configuration are $\beta\simeq 0.772$ and $\gamma\simeq 2.187$. The harmonic
energy scale and the degeneracy factors are is $\hbar\Omega_0=10^{-3}$ eV, 
$\eta_2\simeq 10^{6}, \eta\simeq 3 \times 10^{6}$.} 
\label{pot2}
\end{figure}

Our MLS model is an rf-SQUID operating in the quantum coherence regime 
given by the Hamiltonian  
\begin{equation}
{\cal H}_s/(\hbar\,\Omega_0)=\frac{1}{2}\,[-\partial_z^2+(z-\varphi_{bias})^2]
+\beta\,\cos(\gamma\,z)
\label{squid.1}
\end{equation}
where $\Omega_0=2\pi/\sqrt{LC}$ is the harmonic energy with $L$ being the 
inductance of the SQUID loop and $C$ is 
the effective capacitance of the Josephson 
junction, $\beta=E_{J}/\hbar\Omega_0$ is the dimensionless ratio of 
the Josephson energy $E_J$ to the harmonic energy,   
$\gamma=\hbar\sqrt{\frac{L}{C}}\Bigl(\frac{2\pi}{\Phi_0}\Bigr)^2$ is a 
dimensionless scale parameter, $\varphi_{bias}=2\pi\,\Phi_{bias}/\Phi_0$  is 
the effective bias in the flux which is applicable in a current bias junction, 
and $z=2\pi\Phi/(\gamma\,\Phi_0)$ is the flux ($\Phi$) dependent dimensionless 
dynamical variable
($\Phi_0=h\,c/2e$ is the superconducting flux quantum). 
Truncating the Hilbert space dimension at $N$ in the energy  
eigenbasis, the system Hamiltonian 
in (\ref{squid.1}) is written in the diagonal form 
\begin{equation}
{\cal H}_s=\sum_{n=0}^{N-1}\,E_n(\{\zeta\})\,\vert \{\zeta\} ,n\rangle\,
\langle \{\zeta\} ,n \vert
\label{mss.1}
\end{equation}  
where $\{\zeta\}$ describes the set of system parameters 
$\Omega_0,\beta,\gamma,\varphi_{bias}$ 
where $E_n(\{\zeta\})$, and $\vert \{\zeta\} ,n\rangle$ are respectively 
the parameter dependent 
eigenenergies and eigenvectors of the MLS. This set of parameters is 
sufficiently general to accomodate for all possible effects including the 
degeneracy in the qubit subspace, the symmetry of the wavefunctions etc. 
[we define the degeneracy factor by $\eta=(E_2-E_1)/(E_1-E_0)$ for MLS 
and $\eta_2=(E_1+E_0)/(E_1-E_0)$ for 2LS].   
These three parameters $\Omega_0, \beta, \varphi_{bias}$ control the high energy 
harmonic spectrum, the low energy anharmonic spectrum  and the reflection 
symmetry of the rf-SQUID potential respectively.  
At low energies, a simple numerical diagonalization (\ref{squid.1}) reveals 
that there are 
low lying eigenenergy configurations within the double well regime in which the  
SQUID potential is strongly anharmonic. The parameters of the potential can  
therefore be manipulated to search within this regime for those configurations 
satisfying optimal qubit 
conditions. An interesting case here is to find highly degenerate\cite{pseudo} 
levels corresponding  
to the first two eigenstates for the symmetric  
double well potential (i.e. $\varphi_{bias}=0$). This particular case  
has been extensively examined previously for 2LS using semiclassical methods    
 with an arbitrarily weak tunneling between the wells\cite{LG,SB1}. Another 
configuration that turns out to be important in our calculations is the doubly 
degenerate (DD) configuration for systems with $4 \le N$ in which the first 
four levels are pairwise degenerate with large degeneracy factors. The 
double-well potential 
and energies corresponding to both SD and DD configurations are shown in 
Fig's(\ref{pot1}) and (\ref{pot2}) respectively.  

We numerically find that, the relaxation/dephasing (RD) times for the 
MLS can be controlled by the degeneracy $\eta$. Normally,  
degeneracy is also crucial in controlling the dynamical tunneling rates.  
In our calculations however we directly use the system eigenstates.    
Therefore for the highly degenerate configurations 
bare tunneling between symmetric and anti-symmetric parts of the 
wavefunctions can be neglected to a large extend. This turns 
out to be especially important for quantum computation in the sense that once  
the computation is finished the wavefunction can be maximally localized in 
one of the double wells before any measurement or read-out process.

Although we use the truncated rf-SQUID as the N-level model quantum system to 
study decoherence effects, our treatment is not at the microscopic level. The  
rf-SQUID is shown to be an ideal model to study multilevel effects due to the fact that, the transitional dipole couplings between the low lying energy states 
and the high levels are nonnegligible [see Fig.(\ref{dipole.1})]. Any other 
physical Hamiltonian with 
sufficient number of adjustable parameters as well as nonnegligible dipole 
couplings would be suitable for the calculations presented here.  

In the rest of the paper the harmonic energy $\Omega_0=2\pi/\sqrt{L C}$ is a 
free parameter which is used for scaling energy and time.  

\subsection{Coupling to Noise}
The system-noise interaction is considered to be a 
Caldeira-Leggett type {\it inductive} 
coupling between the SQUID's macroscopic flux coordinate $z$ and the 
environmental {\it flux}-like  
coordinate $\hat{\varphi}_e$ expanded in harmonic environmental modes as 
$\hat{\varphi}_e=
\sum_{k}\,\eta_k\,(b_{-k}^\dagger+b_k)$. The system noise 
interaction is simply 
${\cal H}_{int}=\frac{\alpha}{2}\,z\,\hat{\varphi}_e$ where $\alpha$ is some 
number representing the strength of the inductive coupling ($\alpha$ is to be 
normalized by $\Omega_0$ for a dimensionless coupling).  In the diagonal basis 
$\vert \{\zeta\}, n \rangle$ of the model system the interaction Hamiltonian 
is given by 
\begin{equation}
{\cal H}_{int}=\frac{\alpha}{2}\,\sum_{r,s=0}^{N-1}\,(z)_{r s}\,
\vert s \{\zeta\}\rangle \, \langle r\,\{\zeta\}\vert\, \hat{\varphi}_e ~,
\label{sysnoise.1}
\end{equation}
Here $(z)_{r\,s}=\langle \{\zeta\}\,r\vert z\vert \{\zeta\}\,s\rangle$ 
are the noise induced perturbative {\it dipole} matrix elements of the 
macroscopic system 
coordinate $z$ in the MLS's diagonal basis in (\ref{mss.1}).  
For the model MLS described by (\ref{squid.1}), the dipole matrix elements are  
real and symmetric. 

The rf-SQUID poses a general example in which the multilevelledness of the 
master system manifests itself by finite dipole transition matrix elements  
for both the symmetric (i.e. $\varphi_{bias}=0$) and asymmetric 
(i.e. $\varphi_{bias} \ne 0$) potential configurations. 
In the symmetric contribution  
even parity transitions vanish which results in manifestly off-diagonal 
system-noise 
coupling. Physically, this is in contrast to the most popular models used in the 
literature. On the other hand, when the potential is tilted, the parity 
selection no more holds by which finite diagonal couplings are also created.  

\begin{figure}
\includegraphics[scale=0.5]{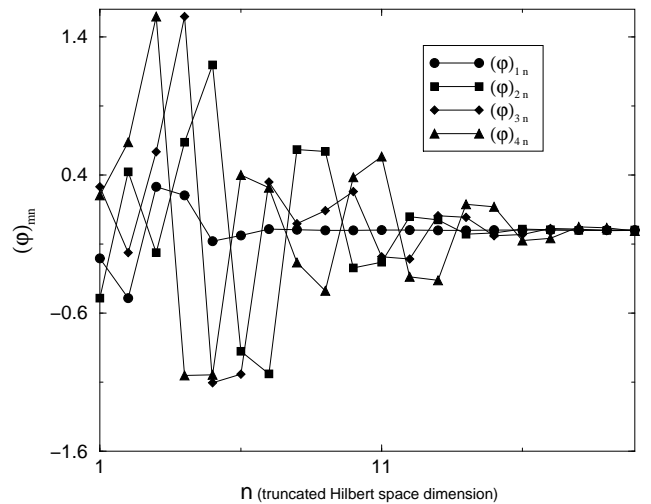}
\caption{A few non-zero dipole matrix elements $(z)_{n,m}$ of the 
coupling of the rf-SQUID to a flux noise versus the truncated Hilbert space 
dimension [calculated in the eigenenergy basis of (\ref{squid.1})]} 
\label{dipole.1}
\end{figure}

In Fig.\ref{dipole.1} the noise induced couplings for an  
asymmetric potential is plotted as a function of the truncated Hilbert space 
dimension.  
Data indicate that the induced dipole strengths between the qubit and the 
higher energy states are 
comparable to those among the qubit states. Therefore, the high energy 
transitions cannot be trivially ignored. The high energy transitions normally  
appear as a result of resonant interactions with the high energy sector of the 
noise spectrum under long interaction times. 
However, in the reduced system these transitions appear in the short time  
dynamics as well, and the short time dynamics is dominated by the nonresonant 
processes. Considering that  
decoherence is dominantly affected by the short time behaviour,   
the nonresonant processes are expected to have observable effects in the 
decoherence properties of the RDM.  
Indeed we observe these effects in the solution of the master equation 
for the MLS (section III.B and C).  

The next is to consider the environmental spectrum and the availability  
of the bath frequencies for these excitations. Regarding this, we consider 
a thermal Gaussian environment spectrum 
\begin{equation}
I(\omega)= 
\omega^{1+\nu}\,e^{-\omega^2/4\Lambda^2}\,\coth(\omega/2T) 
\label{bath.1}
\end{equation}
where $\Lambda$ is the effective noise cutoff frequency and $\nu$ describes 
the subohmic (i.e. $\nu<0$), superohmic (i.e. $\nu>0$), and ohmic (i.e. $\nu=0$) 
character of the spectrum. 
The three environmental parameters $\nu, \Lambda, T$ determine 
the sectors of the spectrum where the system-noise coupling is most effective. 
For $-1\stackrel{<}{\sim}\nu$ (extreme subohmic),  
two regions are of particular importance: 
a) at sufficiently low temperatures and high cutoff corresponding to 
 $T \ll \omega < \Lambda$, the dominant mechanism of relaxation is 
through spontaneous deexcitations\cite{note1}. We call this region 
{\it region}-I;  
b) at high temperatures and high cutoff the region 
$0 \le \omega \ll {\rm min}(\Lambda,T)$ provides a wider range of 
strong environmental couplings which we call as {\it region}-II. If the  
character of the spectrum is more like ohmic or superohmic, i.e. 
$\nu\simeq 0$ or $\nu \simeq 1$ respectively, 
there is lesser room for dexcitations as the availability of the low  
energy modes is suppressed. Therefore, in the ohmic and superohmic regimes,   
the {\it region}-II dominates the RD phenomena.  

\begin{figure}
\includegraphics[scale=0.45]{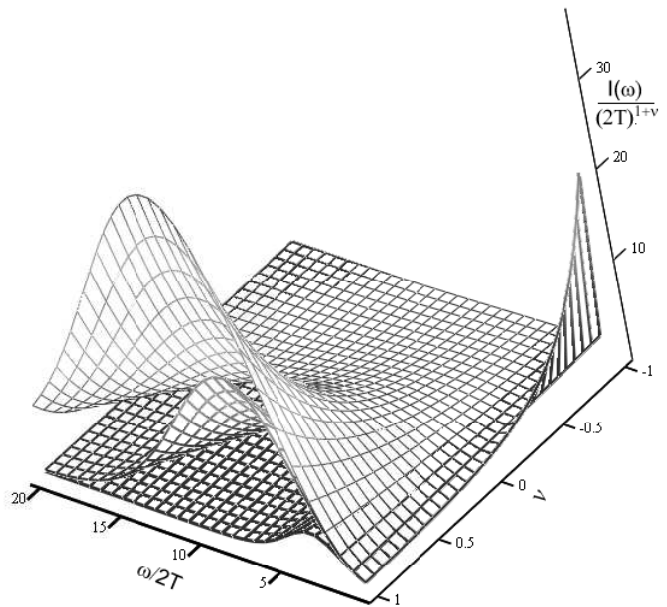}
\caption{The variation of $I(\omega)$ in (\ref{bath.1}) versus $\omega$ and 
$\nu$ for $-1 \le \nu \le 1$ parameterized by $\Lambda/T=10,50,100$ (from the 
innermost to the outermost surfaces respectively).}   
\label{spectrum.1}
\end{figure}

Another feature of (\ref{bath.1}) is related to the majority of   
critical crossover behaviour in the vicinity of $\nu=0$ as predicted earlier by 
Leggett et al.  and depicted in  Fig.\ref{spectrum.1}. 
In this figure, the $\nu \simeq 0$ is a critical vicinity in the Ohmic region   
 separating the subohmic $\nu < 0$ regime from that $0 < \nu$. In the subohmic 
regime $I(\omega)/(2T)^{1+\nu}$ is very small except for vanishingly 
small frequencies ($\omega/2T \ll 1$). Whereas, in the regime $0 < \nu$ 
the maximum value of $I(\omega)$ is observed at higher frequencies  
$\omega \simeq 2\Lambda\sqrt{(1+\nu)/2}$ with an intensity proportional to 
$(\Lambda/T)^{1+\nu}$. 

\section{Master Equation and the Reduced Density Matrix for the MLS}
In the study of decoherence effects due to the weak environmental 
influences, one conventional way is to calculate the time dependent 
RDM elements by solving the master equation. This formalism has been known 
since the independent works of 
Bloch, Redfield and Fano (BRF)\cite{BRF} on spin magnetic resonance and  
widely applied to the spin-boson systems for which many standard references 
exist\cite{appl_BRF}. The standard BRF formalism assumes Markov conditions  
for the solution of the master equations, which leads to exactly solvable   
results for 2LS\cite{TSSexact}. However, the Markovian  assumption is not 
free of drawbacks and that was questioned originally in \cite{AK} and lately 
in \cite{SSO} as well as \cite{GPW} in the context of spin magnetic resonance 
and relaxation. 

In this work, the system noise 
kernel is treated with its most general non-Markovian character\cite{NM}. 
The time evolution of the RDM is obtained in the 
interaction representation by  
\begin{equation}
-i\hbar\,\frac{d}{dt}\hat{\tilde{\rho}}(t)=
[\hat{\tilde{\rho}}(t),\tilde{\cal H}_{int}(t)] 
\label{densmatr.1}
\end{equation}
where $\tilde{}$ denotes the interaction picture. In the context of 
decoherence, we will give more emphasis on short observational times 
in the solution of (\ref{densmatr.1}). A convenient way to proceed is then 
to apply the Born approximation in which the full density matrix is initially  
a product of the system and environmental ones  
(i.e. $\hat{\tilde{\rho}}(0)=\hat{\tilde{\rho}}^{(S)}(0)\otimes
\hat{\tilde{\rho}}^{(n)}(0)$) and at any later and short time approximately 
separates as 
$\hat{\tilde{\rho}}(t)=\hat{\tilde{\rho}}^{(S)}(t)\otimes
\hat{\tilde{\rho}}^{(n)}(0)$.    

The exact iterative solution of (\ref{densmatr.1}) including the second order 
in the interaction 
with the partial trace performed over the environmental degrees of freedom 
yields for the RDM the integro-differential equation 
\begin{equation}
\frac{d}{dt}{\tilde{\rho}}_{nm}^{(S)}(t)=-\int_{0}^{t}\,dt^\prime\,
\sum_{r,s}\,K_{rs}^{nm}(t,t^\prime)\,
\tilde{\rho}_{rs}^{(S)}(t^\prime)
\label{densmatr.2}
\end{equation}
in which we adopt the model interaction Hamiltonian (\ref{sysnoise.1}) for the  
system-noise kernel which is found to be  
\begin{equation}
\begin{array}{lrlr}
&&K_{rs}^{nm}(t,t^\prime) \\
&&=\frac{\alpha^2}{4}\Bigl\{{\cal F}(t-t^\prime)
[(\hat{\tilde{z}}_t\hat{\tilde{z}}_{t^\prime})_{n\,r}
\delta_{s,m}-
(\hat{\tilde{z}}_{t^\prime})_{n\,r}
(\hat{\tilde{z}}_{t})_{s\,m}] \\ 
&&+{\cal F}^*(t-t^\prime)
[(\hat{\tilde{z}}_{t^\prime}\hat{\tilde{z}}_{t})_{m\,s}
\delta_{r,n}-(\hat{\tilde{z}}_{t})_{n\,r}
(\hat{\tilde{z}}_{t^\prime})_{s\,m}]\Bigr\}
\end{array}
\label{densmatr.3}
\end{equation}
Note the the kernel depends on two times as a signature of the non Markovian 
treatment and it is not time translationally invariant. 
Here ${\cal F}(t-t^\prime)={\cal F}^*(t^\prime-t)$ is the complex noise 
correlation function  
\begin{equation}
\begin{array}{lrlr}
{\cal F}(t-t^\prime)&=&{\cal T}r_{n}\Bigl\{\hat{\tilde{\varphi}}^{(n)}(t)\,
\hat{\tilde{\varphi}}^{(n)}(t^\prime)\,\rho^{(n)}(0)\Bigr\} \\
&=&\langle \,
\hat{\tilde{\varphi}}^{(n)}(t)\,\hat{\tilde{\varphi}}^{(n)}(t^\prime)\,\rangle
\end{array}
\label{noisecorrelator.1}
\end{equation}
and 
\begin{equation}
\hat{\tilde{z}}_{t}=\sum_{k,\ell=0}^{N-1}
(z)_{k\ell} e^{i(E_k-E_\ell)t}
\vert k\rangle \, \langle \ell\,\vert
\label{system_field}
\end{equation}
is the time dependent dipole operator in the interaction picture   
where $E_k, \vert k\rangle$  
comprise the eigensolution of the model system. Expanding the noise field 
$\hat{\varphi}_e$ in the independent harmonic modes  
and calculating (\ref{noisecorrelator.1}) in thermal equilibrium  
one obtains the standard thermal noise correlator  
\begin{equation}
{\cal F}(t-t^\prime)=2\sum_k\eta_k^2[\coth(\omega_k/2T)\,
\cos\omega_k(t-t^\prime)-i\sin\omega_k(t-t^\prime)] 
\label{noisecorrelator.2}
\end{equation}
The Markovian versus non Markovian character of the solution of 
(\ref{densmatr.2}) is 
determined in the weak system-noise interaction limit by the competition of 
three time scales: $\tau_B$, noise correlation time scale, $\tau_{R}$ and   
 $\tau_{dep}$, the relaxation and dephasing time scales\cite{BP} 
of the reduced system respectively. The noise correlation time scale   
 is found roughly from by the thermal Gaussian bath spectral width as 
$\tau_B \simeq 1/\Lambda$. 
At the Markovian limit, the environmental correlation time 
$\tau_B\simeq \Lambda^{-1}$ is much smaller than the RD times.   
For two level systems this condition can be  
met provided that the system noise-coupling is sufficiently small. 
However, for MLS, the question of whether the Markovian condition is 
satisfied is more nontrivial. The basic reason is that in the MLS there   
is a larger number of  time scales and decay channels of which presence 
may considerably reduce the effective decoherence times.  

In this work, the numerical solution of (\ref{densmatr.2}) 
is performed by discretizing time in steps $\Delta t=10^{-2}\,\Lambda^{-1}$.   
The Hermiticity and 
the normalization of the RDM at each time step is maintained within an 
accuracy of $10^{-25}$.  

\subsection{The system-noise Kernel} 
In the model Hamiltonian defined in (\ref{squid.1}) all energies and time scales 
(particularly the RD times) are given in units of 
$\hbar \Omega_0$ and $\Omega_{0}^{-1}$ respectively.  
The parameters affecting the numerical results are, 
 $\nu, \Lambda, T$ for the thermal noise, $\alpha$ for the system-noise bare 
coupling and the dipole matrix elements $(z)_{n\,m}$ for the pure MLS.   
The noise spectrum is assumed to be continuous of which the real part is 
responsible for RD effects  
and is given by the spectral density in (\ref{bath.1}). In 
the Markovian limit the imaginary part of ${\cal F}(t-t^\prime)$ is 
vanishingly small and the resulting Lamb-type energy renormalization effects 
are 
negligible. In the numerical calculations however we include the full complex 
noise correlation function as\cite{HPZ}

\begin{equation}
\begin{array}{lrlr}
{\cal F}(t-t^\prime)=&&2\int_{0}^{\infty} d\omega\,  
\omega^{1+\nu}e^{-\omega^2/\Lambda^2} \\
&& \times \Bigl\{\coth(\omega/2T)
\cos{\omega(t-t^\prime)} 
-i\sin{\omega\,(t-t^\prime)}\Bigr\}
\end{array}
\label{realcorrelator.1}
\end{equation}

Inserting (\ref{realcorrelator.1}) in (\ref{densmatr.3}) we obtain the 
system-noise kernel for our model. An upper frequency cutoff of 
$\omega_{max}=5\Lambda$ is used in the numerical integral in 
(\ref{realcorrelator.1}).  

\subsection{Overview of the 2LS results}
\begin{figure}
\includegraphics[scale=0.45]{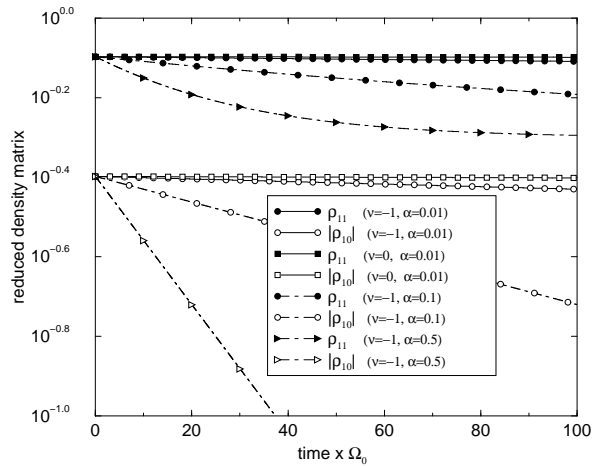}
\caption{Time dependence of the RDM in units of $\Omega_0^{-1}$ 
for various representative $\alpha,\nu$ parameter sets at $T=0$ and 
$\Lambda=0.1$. For the model system   
the potential is symmetric and the bare TLS is in SD configuration.}
\label{fig4}
\end{figure}

We now examine the time behaviour of the RDM in (\ref{densmatr.2}) for a 2LS. 
The solution is shown in Fig.\ref{fig4} on the logarithmic scale  
 for a few representative parameters and for the SD configuration.
The degeneracy parameter is $\eta_2 \sim 10^6$. 
We also fixed $\alpha=0.01$ in the rest of the work unless otherwise stated. 

In Fig. (\ref{fig4}) the first observation is that, exponential 
RD is effective immediately in the short time 
regime $t < 20 \Omega_{0}^{-1}$. We also confirmed numerically that the 
asympthotic time behaviour as well as the RD rates 
are independent of the initially prepared state. As the asymptotic time 
behaviour is concerned, for symmetric   
configurations (pure $\sigma_x$ coupling), the density matrix  
converges to the maximum entropy (informationless) limit $\hat{I}/2$,   
where $\hat{I}$ is the unit matrix, irrespectively from the spectral properties 
of the noise or the system-noise coupling. The results also indicate that   
the relaxation time scale $\tau_R$ (read from the filled symbols) 
and the dephasing time scale $\tau_{dep}$ (read from the hollow symbols) are 
compatible. This result is in agreement particularly with the recent exact 2LS 
calculations 
using the path integral influence functional formalism\cite{SB2}. 

\begin{figure}
\includegraphics[scale=0.45]{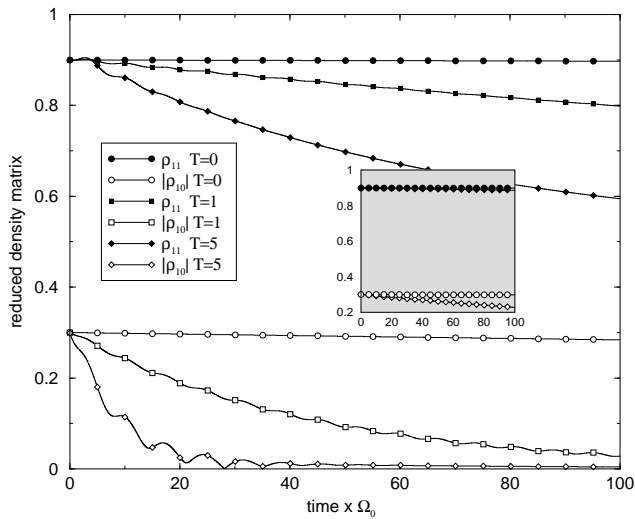}
\caption{Time dependence of the 
RDM when the system potential is biased by $\varphi_{bias}=0.2$. 
The other parameters are $\alpha=0.01, \Lambda=0.1$ and $\nu=-1$. 
Comparing this figure 
with the previous one, a crossover in the time dependence of the RDM is
observed in the $\nu-T$ plane. The $\nu=0$ data is also added in the inlet 
to facilitate this comparison.} 
\label{fig4b}
\end{figure}

\begin{figure}
\includegraphics[scale=0.45]{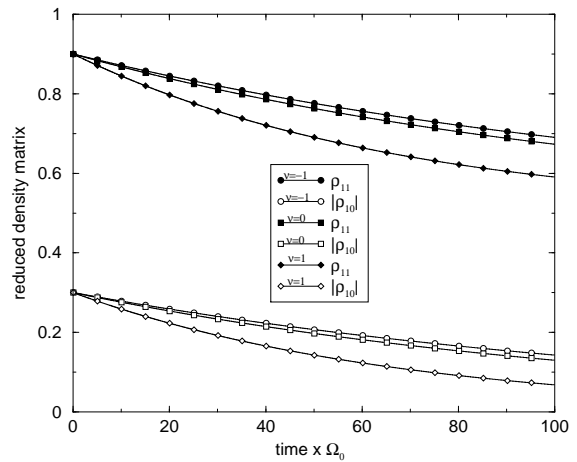}
\caption{Time dependence of the RDM in the SD configutation. The other 
parameters are fixed at $T=0$, $\alpha=0.01$ and $\Lambda=1$.}  
\label{fig5}
\end{figure}

\begin{figure}
\includegraphics[scale=0.45]{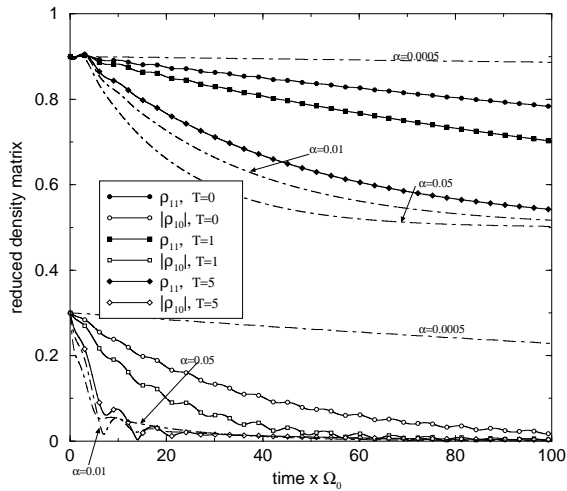}
\caption{The $\alpha-T$ behaviour of the RDM is shown for an asymmetric 
potential with $\varphi_{bias}=0.2$. The 
noise parameters are $\alpha=10^{-2}$, $\Lambda=5$ and $\nu=-1$.}
\label{alpha.T}
\end{figure}

\begin{figure}
\includegraphics[scale=0.45]{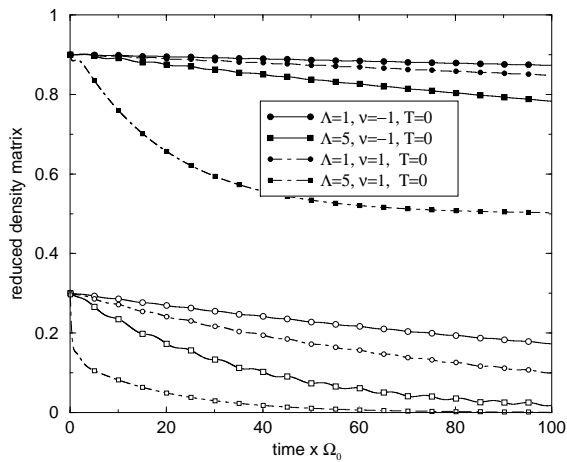}
\caption{The density matrix parameterized by different $\Lambda$ and $\nu$ 
values at $T=0$ and $\alpha=0.01$  
for the asymmetric potential with $\varphi_{bias}=0.2$. 
The crossover from the weakly oscillating damped behaviour 
at short times to strong relaxation at longer times can be observed.}
\label{alpha.nu}
\end{figure}

We also confirm that all regions in the noise spectrum have strong 
influence on RD. For this observation, one has to compare 
Fig's. \ref{fig4}-\ref{alpha.nu} correponding to different spectral 
properties and temperatures. For instance, for  
$\alpha=0.01, \Lambda=0.1$ and $T=0$ [see Fig.\ref{fig4}],   
we recover the under damped and weak dephasing 
limit of \cite{SB1} for all $\nu$. In addition, no oscillations are 
observed in the SD configuration. 

Larger RD rates are observed as $\nu$ is made to be more negative towards 
$\nu=-1$. We identify this behaviour as the manifestation of the region-II in  
the noise spectrum (see the end of section II.A). More data are also shown 
in the same figure for indicating the effect of various $\alpha$ 
values. In Fig.\ref{fig4b} the nondegenerate case with an 
energy difference between the levels $\Delta E=1.2 \hbar \Omega_0$ is shown  
for $\alpha=0.01, \nu=-1, \Lambda=0.1$ and for various temperatures.  
The $\nu=0$ and $\nu=1$ curves again yield weaker RD rates  
within the indicated temperature ranges. Another noticable feature in  
Fig.\ref{fig4b} is the weakly oscillating behaviour. The weak oscillations  
are more prominent at high temperatures and short times. In this case 
for $\omega \ll T$ the system-noise coupling is large due to   
the large number of thermally activated environmental modes. 
  This behaviour, which is characterized by weakly damped 
Rabi-like oscillations, was predicted in the analytic calculations of Leggett 
 et al. In order to examine the influence of the spectral width $\Lambda$    
a similar zero temperature plot as in Fig.\ref{fig4} is made in Fig.\ref{fig5} 
 for a wider spectral width $\Lambda=1$. 
Comparing this figure with Fig.\ref{fig4}, a crossover in the time dependence 
of the RDM can be observed in the $\Lambda-\nu$ plane.   
(The crossover can also be activated thermally as to be seen in the following 
Fig.'s \ref{alpha.T} and \ref{alpha.nu}.) 

There are a number of ways to increase the effective system-noise coupling.  
The following Fig.\ref{alpha.T} gives a sample
from the $\alpha-T$ behaviour for the asymmetric potential configuration.
Increasing the temperature increases the coupling by
filling the available photon modes for $\omega < T$. The sample 
data is shown for $T=0,1,5$ at $\alpha=0.01, \Lambda=5$ and $\nu=-1$ 
(indicated  
by the filled and opaque symbols connected with solid lines). The second way to 
increase the system-noise coupling is to   
directly increase $\alpha$ (indicated by the dotted dashed lines). In the small 
coupling regime, for which the sample data is shown for 
$\alpha=5\times 10^{-4},10^{-2},5 \times 10^{-2}$ at $T=0$ $\nu=-1$ and for an 
increased $\Lambda$ ($\Lambda=5$),  
the RDM experiences stronger damping. 
The weakly damped oscillations survive at short times at finite and small 
temperatures. The larger the temperature the larger the amplitude of the
oscillations and the faster they diminish.

For completeness we also add in Fig.\ref{alpha.nu}
the behaviour in the $\Lambda-\nu$ plane at zero temperature. A comparison  
between the Fig's \ref{alpha.T} and \ref{alpha.nu} reveals a temperature
modulated crossover (confirm, for instance, a similar decay of the sets for
$\Lambda=5, \nu=-1$ at $T=5$ in Fig.\ref{alpha.T} with
$\Lambda=5, \nu=1$ at $T=0$  in Fig.\ref{alpha.nu}).

The  major difference of the model interaction Hamiltonian in 
(\ref{sysnoise.1}) from the standard ($\sigma_z$-type) spin-boson model  
is in the manipulation 
of the potential. In contrast to the standard spin-boson model, 
in our case only non-diagonal, $\sigma_x$, type coupling is present under  
the symmetric potential [see Fig.\ref{dipole.1}]. As a result, dramatic 
differences in the time dependence of the reduced system are observed between 
the two models. For instance, the diagonal coupling is standardly considered 
for the study of pure dephasing. In this type of coupling the relaxation is 
manifestly forbidden and the initial states do not change their populations. 
The diagonal coupling also yields strongly temperature dependendent dephasing 
rates with the rates vanishing at $T=0$. 
On the other hand, when the system-noise coupling is not diagonal, the induced 
transitions between the system states can probe the entire noise spectrum 
creating decoherence even at zero temperature. These induced transitions are 
nonresonant and they have observable effects particularly in the short time 
dynamics of the RDM\cite{Bruder}. The RD times observed as 
the result of such system-noise coupling are expected to be nonzero even at zero 
temperature. This characteristic behaviour of the non-diagonal coupling is 
confirmed in our calculations both for the 2LS in Figs.\ref{fig4}-\ref{alpha.nu} 
and for the MLS in the following sections. Recently, there are other claims 
using 
realistic models on decoherence effects in mesoscopic systems\cite{Golub} as 
well as a few experimental confirmations on the 
saturation of the RD rates at low temperatures\cite{Webb}. 

A curious observation in Fig's \ref{fig4}-\ref{alpha.nu} is the strong 
dependence of RD  
time scales on the spectral width $\Lambda$. A naive expectation
is that for $\Delta E \ll \Lambda$ and at very small temperatures the resonant
transitions are unfavoured and there are no environmental modes available
therefore the relaxation should be inhibited. The point that is often 
missed in this popular 
argument is the different role played by the short time nonresonant transitions. 
The resonant transitions are favoured when the system
interacts with the environment at {\it sufficiently large times}.
The system however relaxes
differently at short times by prefering to stay off-resonant in its
interaction with
the noise field thereby sampling all regions of the noise spectrum. This 
causes the strong dependence on $\Lambda$ we observe at short times.

In summary, it is confirmed that the rich transient effects are observed 
usually in the short time behaviour 
in which all energy scales in the noise spectrum take part rendering the 
relaxation process sensitive to the relative magnitudes of those scales. 
We confirmed the several crossover regions that have 
been predicted in the path integral influence functional calculations. 

The decoherence and dephasing dynamics is governed by all frequency regions 
in the noise spectrum. In particular, the short time behaviour is affected 
strongly by all frequency regions due to the nonresonant processes. 
For $\Lambda \ll 1$, Fig.\ref{fig4} indicates 
that as $\nu$ increases in the interval $-1 \le \nu \le 1$ the relaxation rates 
gradually increase still remaining in the weak relaxation regime. A crossover in the 
$\nu-\Lambda$ plane is observed [compare with Fig.\ref{fig5}] as $\Lambda$ 
is increased. With this being the case for symmetric potentials, for asymmetric 
 ones the additional feature of weak oscillations are present 
in the short time dynamics.

\subsection{MLS with $3 \le N$}

The effect of multileveledness on decoherence has not yet received 
the attention that it deserves 
in the literature. This is, in part, due to the lack of practical analytic   
tools in solving the master equation. The complexity of the formal methods 
such as   
the noninteracting blip approximation increases at each time step as $N^2$ 
which renders the analytic sum over all virtual configurations in the path 
integral approach intractable. Usually, the common argument is that, for 
sufficiently low temperatures, a multilevel system, of which the first two levels 
(the qubit) are sufficiently well separated from the rest, behaves as a two 
state system. We have already observed that there are two major pitfalls in 
this assumption. Firstly, it excludes the very realistic case in which 
decohering effects are induced through interactions   
with a strongly fluctuating quantum field. In such a case, the fluctuations  
in the distribution of environmental modes in the noise spectrum is the major  
source of decoherence. 
 Secondly, and more generally, the short time 
behaviour of a MLS -which is the most prominent regime in 
the quantum computational perspective- is affected by a large frequency region  
in the noise spectrum. These comprise the basic motivation   
of why we should look into the effect of higher levels on decoherence in a 
MLS. We will continue by examining the cases $N=3$ and $3 < N$ separately. In 
particular, we remark on certain interesting decoherence properties of 
the doubly degenerate systems with $4 \le N$.  

\begin{center}
{\bf $N=3$ case}
\end{center}

\begin{figure}
\includegraphics[scale=0.45]{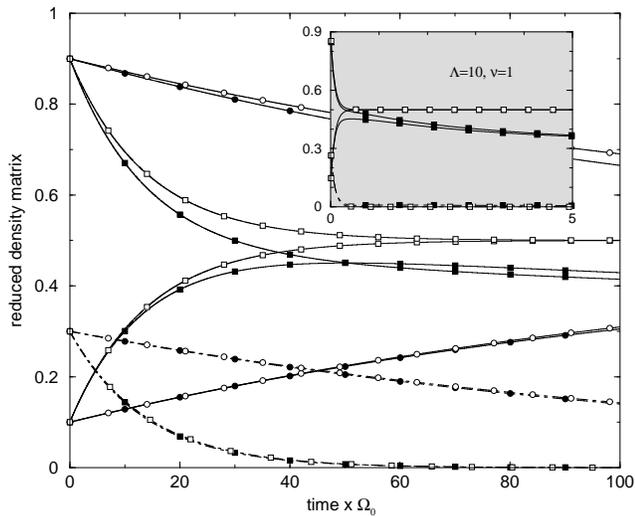}
\caption{Comparison of the effect of the spectral width $\Lambda$  
on the time dependence of the elements $\rho_{00}, \rho_{11}$ and 
$\vert\rho_{01}\vert$ between the 2LS and 3LS. The fixed parameters are 
$T=0$, $\alpha=0.01$ and $\nu=-1$. The open symbols refer to the 
case $N=2$ and the solids ones refer to $N=3$. Also the solid lines are the 
diagonal elements $\rho_{00}$ (with $\rho_{00}(0)=0.1$) and 
$\rho_{11}$ (with $\rho_{11}(0)=0.9$), the dotted-dashed lines are 
the non-diagonal ones $\vert \rho_{10}\vert$ 
(with $\vert \rho_{10}(0)\vert=0.3$). More specifically, circle is 
$\Lambda=1$, square is $\Lambda=10$. The inlet is the case $\nu=1$ and  
only $\Lambda=10$ is displayed for simplicity.} 
\label{3lsa}
\end{figure}

From the quantum computational point of view, the three level systems 
are as important as the two level ones (see for instance \cite{Kulik}). 
Recently, the Linblad approach was used for the RDM of a 
multilevel system in connection with the quantum Zeno effect\cite{3LS_zeno}. 
The Linblad equation is derived directly from the Bloch-Redfield-Fano 
equation for the RDM and is based on the validity of the  
Markovian condition $\Lambda^{-1} \ll \tau_R$. 
We compare in Fig.\ref{3lsa} 
  the solutions of the RDM for $N=2$ and $N=3$  
 at $\alpha=0.01$, $T=0$ and for    
$\Lambda=1, 10$. We look at the symmetric potential in the SD configuration.  
For 3LS, the observed energies form a $\Lambda$-shaped configuration
and, in units of $\Omega_0$, are roughly $E_0\simeq E_1 \approx 0.1$ and
$E_2 \approx 2.3$. The degeneracy parameter $\eta=(E_2-E_1)/(E_1-E_0)
\simeq 10^6$ and the third level $E_2$ is above the double well barrier 
as shown in Fig.\,(\ref{pot1}).  
Here we mainly discuss the subohmic case $\nu=-1$. 

When $\Lambda \ll 1$ the resonant coupling of the
first two levels to the third level is very weak. In principle, at short 
observational  
times, the nonresonant excitations are induced with frequencies much higher than 
the resonant frequency $E_3-E_2 \simeq 2$ (in units of $\Omega_0$). When 
$\Lambda \ll 1$ however, these transitions are suppressed by the Gaussian 
cutoff. As a result, the 3LS is  basically confined to its highly degenerate 
qubit subspace. This confinement can be observed all the  
way up to much higher spectral widths such as $\Lambda=1$ as long as the 
Gaussian suppression is manifested. This behaviour is shown in Fig.\ref{3lsa} 
for $\Lambda=1$ and $\Lambda=10$.   
For considerably long duration (i.e. $\sim 100\times \Omega^{-1}_0$) and 
for $\Lambda=1$ the qubit subspace in the three level system   
has a negligible leakage into the third level. 
For much larger $\Lambda$ such as $\Lambda=10$, the third level is allowed 
to participate in the transitions. The RD   
rates are therefore found to be significantly larger 
than that of the $N=2$ case before. 

An enhancement in the high frequency and suppression in the low frequency 
coupling as compared to $\nu=-1$, is generated if we now consider $\nu=1$ 
(shown in the inlet of Fig.\ref{3lsa}). Under these conditions, the short time 
behaviour is dramatically   
influenced by a strong leakage to the third level (note the short 
time span on the horizontal axis between zero and $5\times \Omega_0^{-1}$).  

The asymptotically long time dynamics is independent of  
the system-noise parameters. In sufficiently long time the system looses  
all the information that is put in the initial state: 
$\rho_{kk}(t\to \infty)=1/3, (k=0,1,2)$ and 
$\vert \rho_{k\,j}(t\to \infty)\vert=0, (k\ne j)$. 

\begin{figure}
\includegraphics[scale=0.4]{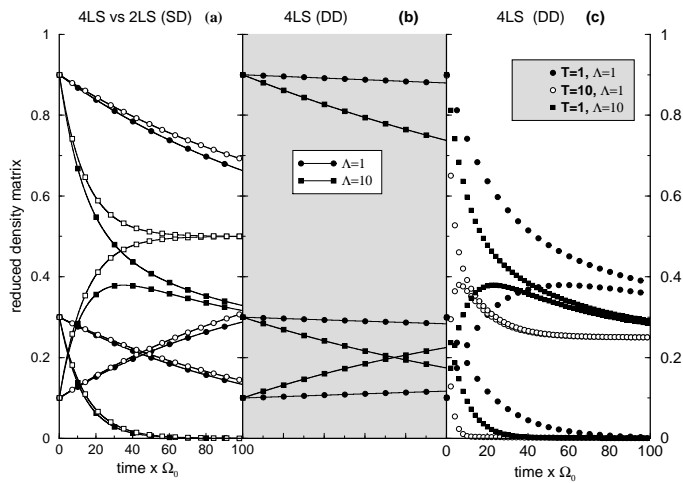}
\caption{(a) Comparison, at $T=0$ and $\nu=-1$, of the effect of the spectral 
width $\Lambda$  
on the time dependence of the elements $\rho_{00}, \rho_{11}$ and 
$\vert\rho_{01}\vert$ between the 2LS and 4LS. The symbols are the same 
as in Fig.\ref{3lsa}. The figure describes the singly degenerate case 
$\eta\simeq 10^6$; (b) the rates at $T=0$ and $\nu=-1$ for the doubly 
degenerate (denoted by DD in the figure title)  
configuration $\eta_2=(E_2-E_1)/E_1-E_0)\simeq 3\times 10^6$ and 
$\eta_1=(E_3-E_2)/E_2-E_1)\simeq 10^6$ [horizontal and vertical axes have 
the same span as (a)]; (c) the thermal case at the indicated $\Lambda, T$ 
values at $\nu=-1$.}  
\label{fig4ls}
\end{figure}

\begin{center}
{\bf $N=4$ case}
\end{center}
We compare the 4LS with a 2LS in Fig.\ref{fig4ls} for 
$\nu=-1$. Here, we 
have three sets of curves indicated by (a), (b) and (c). In Fig.\ref{fig4ls}.a 
the 4LS is compared  to 2LS when both systems are highly degenerate. 
In  Fig.\ref{fig4ls}.b the four level system is doubly degenerate and at zero 
temperature. The third set of curves  
are plotted in Fig.\ref{fig4ls}.c corresponding to the DD configuration at 
finite temperatures. 

For the singly degenerate (SD) case, the qualitative features  
between the 2LS and the 4LS are similar to the previously discussed 
case between 2LS and the 3LS. Here, we observe for the diagonal elements,  
a much higher relaxation rate (as well as leakage) out of the qubit subspace 
during the observed   
time although the dephasing rates are indistinguishable for the 2LS and the 
4LS. The RDM asymptotically approaches to the informationless limit 
$\hat{\rho}=\hat{\bf I}/4$. 

For $N=4$ we have the chance to look at the DD configuration as depicted in  
Fig.\,(\ref{pot2}). For the doubly degenerate configuration, the degeneracies 
are as high as $\eta_2=(E_2-E_1)/E_1-E_0)\simeq 3\times 10^6$ and 
$\eta_1=(E_3-E_2)/E_2-E_1)\simeq 10^6$. We surprisingly observe a 
tremendous suppression (by almost two orders of magnitude) at 
sufficiently small temperatures     
in the RD rates (Fig.\ref{fig4ls}.b). The rates 
and the DD-suppression strongly depend on the temperature. 
A comparison of the $T=0, \Lambda=1$ in Fig.\ref{fig4ls}.b and $T=1, \Lambda=1$
in Fig.\ref{fig4ls}.c can reveal this strong dependence.   

\begin{figure}
\includegraphics[scale=0.45]{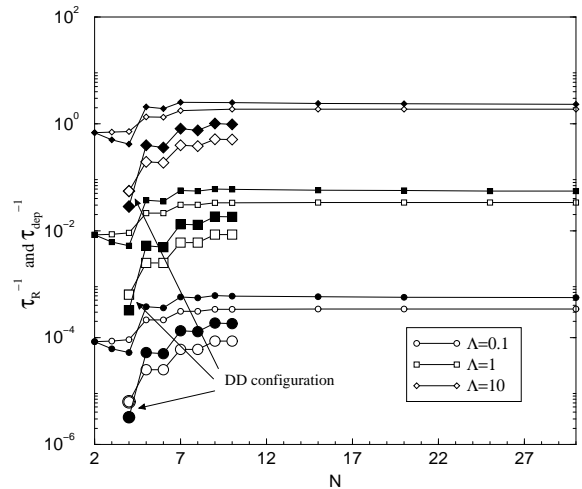}
 \caption{Relaxation and dephasing rates against the number of levels 
for different spectral widths at $T=0$ and $\nu=0$. Note the logarithmic 
vertical axis. Small symbols refer to the singly degenerate MLS and the 
larger symbols refer to doubly degenerate one. The open and solid symbols  
refer to dephasing and relaxation times respectively.}
\label{relax_n}
\end{figure}

\begin{center}
{\bf $4 < N$ case}
\end{center}
In order to extract some quantitative numbers for the RD times 
we made use of the numerical observation that for weak system-environment  
coupling the time dependence is approximately exponential  
at short times. We then follow \cite{SB1} and write for the time dependence   
of a general matrix element at short times 
\begin{equation}
\vert\rho_{ij}(t)\vert\simeq \vert\rho_{ij}(\infty)\vert+[\vert\rho_{ij}(0)\vert-
\vert\rho_{ij}(\infty)\vert]exp(-t/\tau_{ij})~.  
\label{relax_time.1}
\end{equation}
The RD times are extracted from the time dependence of 
$\rho_{11}(t)$ and $\vert \rho_{10}(t)\vert$ respectively as 
\begin{equation}
\tau^{-1}_{ij}\simeq -\frac{1}{1-\vert\rho_{ij}(\infty)/\rho_{ij}(0)\vert}
\frac{d\ln\vert\rho_{ij}\vert}{d t}\Bigl\vert_{t=0}
\label{relax_time.2}
\end{equation}
where $i=j=1$ is used in the calculation of the relaxation rate  
 and $i=0,j=1$  
is for the qubit dephasing rate corresponding to the first excited level. 
For the RDM at asympthotic times we have $\rho_{11}(\infty)=1/N$ and   
$\vert\rho_{10}(\infty)\vert=0$. The equation (\ref{relax_time.2})  
breaks down when $\vert\rho_{ij}(0)\vert=\vert\rho_{ij}(\infty)\vert$ 
which we stay away 
by appropriately choosing $\rho_{ij}(0)$. In Fig.\ref{relax_n} 
the data are represented at zero temperature and   
$\nu=0$. To be used in (\ref{relax_time.2}), the initial conditions 
are set at $\rho_{00}(0)=0.2, \rho_{11}(0)=0.8, \rho_{10}(0)=0.4 i$ with all 
density matrix elements outside the qubit subspace zero. 
Three different curves stand for (bottom to top) $\Lambda=0.1,1,10$ 
with the open symbols corresponding to dephasing and the solid ones to 
the relaxation rates. Each set of data is shown for SD as well as DD  
configurations separately. Also note that the vertical axis is logarithmic. 

Let us concentrate first on the SD configurations in Fig.\ref{relax_n}. 
In a large $\Lambda$ range $N=4$ and $N\sim 10$ 
appear to be two crucial points. For $4 < N$ relaxation is 
approximately twice faster than dephasing 
and both rates rapidly saturate near $N\simeq 10$ and they are 
independent of $N$ for $10 < N$. The onset of saturation  
is naturally model dependent. In our case this onset coincides with the range  
of strong dipole transition matrix elements of the model in (\ref{squid.1}) 
 (see Fig.\ref{dipole.1}).   
Turning to the DD configurations, 
we observe that for the same environmental parameters and for all $N$, 
decoherence rates 
for the DD case are strongly suppressed by nearly two orders of magnitude  
as compared to the SD configuration.  

With this section we conclude the RD calculations for the interaction   
between the system and the thermally equilibrated noise. 
We now focus our attention on the 
calculations of the transition rates by a different approach, 
the Fermi Golden rule.  

\section{Fermi Golden Rule}
The Fermi Golden Rule (FGR) 
provides a simple and qualitative tool to reproduce many of 
the features of the relaxation times that we observe  in Fig.\ref{relax_n}. 
Quantitative agreement should not be expected between the Fig.\ref{relax_n} 
and the FGR results. This is mainly due to the fact that 
the data produced in  Fig.\ref{relax_n} reflects the effects of the short time 
dynamics whereas, the FGR gives more accurate results for the long time 
resonant 
interactions. In comparing the time scales found by directly solving the RDM 
and by FGR, we keep the absolute 
time scales arbitrary and only compare the qualitative behaviour. 

We assume that the MLS is prepared at $t=0$ in the first excited state 
$\vert\{\zeta\}, 1\rangle$. 
The probability that the system stays in the same state after 
interacting with the environment for a duration $t$ is\cite{SB1}
\begin{equation}
p_{\psi}(t)=\Bigl\vert\langle \psi(0)\vert exp[-\frac{i}{\hbar}
\int_{0}^{t}dt^\prime\tilde{\cal H}_{int}(t^\prime)]
\vert \psi(0)\rangle\Bigr\vert^2 
\label{FGR.1}
\end{equation}
where for our case $\vert \psi(0)\rangle=\vert \{\zeta\}, 1\rangle$.
Including second order perturbation in the dipole couplings with  
an environment in thermal equilibrium, (\ref{FGR.1}) can be written as 
\begin{eqnarray}
&&1-p_{\psi}(t)= \nonumber \\
&&2(\frac{\alpha}{2})^2\sum_{s}\varphi_{sn_0}^2\,\int_{0}^{t}dt^\prime
\int_{0}^{t^\prime}dt^{\prime\prime}e^{i(E_{n_0}-E_s)(t^\prime-t^{\prime\prime})}
{\cal F}(t^\prime-t^{\prime\prime}) \nonumber \\
\label{FGR.2}
\end{eqnarray}
where ${\cal F}(t^\prime-t^{\prime\prime})$ is the environmental 
correlation 
function given by (\ref{noisecorrelator.2}). The relaxation rate 
$\tau_{FGR}^{-1}$ corresponding to the first excited state is found 
by the same method that is described in (\cite{SB1}) 
[The Eq.\,(3.37) therein]. 

\begin{figure}
\includegraphics[scale=0.45]{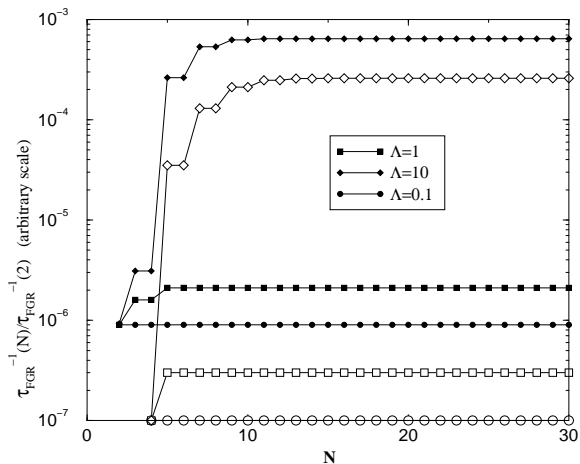}
 \caption{Relaxation rates against the number of levels for the 
SD (solid symbols) and DD (open symbols) cases.}   
\label{fgrmain}
\end{figure}

\begin{figure}
\includegraphics[scale=0.45]{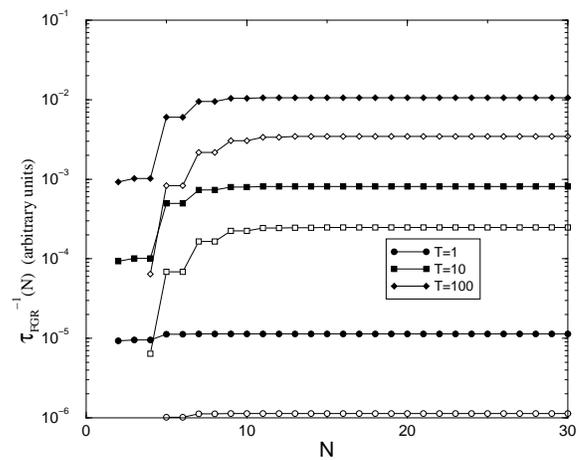}
 \caption{Comparison of the absolute relaxation rates (in common arbitrary 
units) 
with $N$ between the SD (solid symbols) and DD (open symbols) configurations. 
Here, $\nu=0$ and $\Lambda=10$. 
Same temperatures are implied for each SD-DD pair.}   
\label{fig.fgr3}
\end{figure}

The relaxation times found by the FGR are summarized in 
Fig's\ref{fgrmain}, and \ref{fig.fgr3}. 
In Fig.\ref{fgrmain} the relaxation times are plotted against the same 
parameters as in Fig.\ref{relax_n} with the same symbols. The FGR data 
reproduce many of the features in Fig.\ref{relax_n}.  
The first observation is the same offset at $N=4$ and the
saturation of the rates slightly above this offset with a rapid increase
for $4 < N$. The second observation is that by 
increasing the spectral 
width, the relaxation rates can be increased by as much as an order of 
magnitude. 

When the temperature is varied, we observe the same trend as in the previous  
figure as depicted in Fig.\,(\ref{fig.fgr3}). 

\section{Conclusions}
We examined the RD properties of a multilevel system 
using non-Markovian master equation formalism. It is shown that the short time  
behaviour of the 
density matrix is influenced by nonresonant transitions in the MLS receiving 
contributions from all frequency regions in the noise spectrum. For the model 
interaction used, the dipole transitions are nonzero within a finite range of 
 levels. The RD times calculated within this model show a 
saturation within the same range largely independent from the 
system-noise parameters.    

It is generally found that the decoherence effects in MLS are more 
pronounced than those in the 2LS. We observe that a distinct counter example 
is posed by the doubly degenerate MLS with $4 \le N$. The RD    
rates are found to be highly suppressed in comparison with the singly degenerate 
or non-degenerate systems for the same system and environment parameters. These 
result were also confirmed using the transition 
rates found from the Fermi-Golden rule. At the first glance, this curious 
suppression of decoherence reminds us the decoherence free subspaces. 
Nevertheless, the arbitrariness of the parameters of the DD model, and in 
particular of the dipole matrix elements rules out the possibility whether any 
set of invariant states can form a decoherence free subspace under the coupling 
with the environment. In order to understand the true nature 
of this strong suppression, more formal and analytic methods must be 
developed for the DD systems\cite{THEMKS}. 

\section{Acknowledgements}
This research is supported by the Scientific and Technical Research Council of 
Turkey (T\"{U}B{\.I}TAK) grant number TBAG-2111 (101T136). The authors thank   
I O Kulik and E Mese for comments.


\begin{thebibliography}{99}
\bibitem{note1} We assume that the system is initially prepared  
at a given superposition state 
$\psi(0)\rangle=a\vert 0\rangle+b\vert1\rangle$ where $\vert 0\rangle$ and 
$\vert 1\rangle$ are respectively the ground and the first excited states in 
the qubit subspace.
\bibitem{NM} The Markovian versus non Markovian character of the system-noise
kernel and the Markovian versus non Markovian character of the solution of the
RDM should be clearly distinguished. In the former we take the most general
form of the kernel as it comes out of the interacting model which depends on
two times independently. The Markovian or non Markovian character of the
solution however depends on whether the conditions $\tau_{B} \ll \tau_R$
and $\tau_{B} \ll \tau_{dep}$ are satisfied where 
$\tau_{B}, \tau_R, \tau_{dep}$ are the noise correlation, relaxation and 
 dephasing time scales respectively. A good example of Markovian
solution using a non Markovian kernel can be found in the last work in
\cite{SB1} above.
\end{thebibliography}
\end{document}